\documentstyle[12pt]{article}

\font\twlgot =eufm10 scaled \magstep1
\font\egtgot =eufm8
\font\sevgot =eufm7

\font\twlmsb =msbm10 scaled \magstep1
\font\egtmsb =msbm8
\font\sevmsb =msbm7

\newfam\gotfam
\def\pgot{\fam\gotfam\twlgot}
\textfont\gotfam\twlgot
\scriptfont\gotfam\egtgot
\scriptscriptfont\gotfam\sevgot
\def\got{\protect\pgot}

\newfam\msbfam
\textfont\msbfam\twlmsb
\scriptfont\msbfam\egtmsb
\scriptscriptfont\msbfam\sevmsb
\def\Bbb{\protect\pBbb}
\def\pBbb{\relax\ifmmode\expandafter\Bb\else\typeout{You cann't use
Bbb in text mode}\fi}
\def\Bb #1{{\fam\msbfam\relax#1}}

\def\thebibliography#1{\bigskip\section*{REFERENCES}\bigskip\list
{}{\settowidth\labelwidth{#1}\leftmargin\labelwidth
\advance\leftmargin\labelsep
\usecounter{enumi}}
\sloppy\clubpenalty4000\widowpenalty4000
\sfcode`\.=1000\relax}

\def\op#1{\mathop{\fam0 #1}\limits}

\newcommand{\id}{{\rm Id\,}}

\newcommand{\Ker}{{\rm Ker\,}}
\newcommand{\nm}[1]{\mid {#1}\mid}

\newcommand{\beq}{\begin{equation}}
\newcommand{\eeq}{\end{equation}}
\newcommand{\ben}{\begin{eqnarray}}
\newcommand{\een}{\end{eqnarray}}
\newcommand{\be}{\begin{eqnarray*}}
\newcommand{\ee}{\end{eqnarray*}}
\newcommand{\bea}{\begin{eqalph}}
\newcommand{\eea}{\end{eqalph}}

\newcommand{\cG}{{\got g}}

\newcommand{\cP}{{\cal P}}
\newcommand{\cR}{{\cal R}}
\newcommand{\cL}{{\cal L}}
\newcommand{\cV}{{\cal V}}

\newcommand{\cH}{{\cal H}}
\newcommand{\cF}{{\cal F}}
\newcommand{\cS}{{\cal S}}

\newcommand{\cN}{{\cal N}}

\newcommand{\bL}{{\bf L}}

\newcommand{\al}{\alpha}
\newcommand{\bt}{\beta}
\newcommand{\dl}{\delta}
\newcommand{\la}{\lambda}
\newcommand{\La}{\Lambda}

\newcommand{\om}{\omega}
\newcommand{\Om}{\Omega}
\newcommand{\m}{\mu}
\newcommand{\n}{\nu}

\newcommand{\G}{\Gamma}

\newcommand{\ve}{\varepsilon}

\newcommand{\vt}{\vartheta}
\newcommand{\si}{\sigma}
\newcommand{\w}{\wedge}

\newcommand{\wh}{\widehat}
\newcommand{\ol}{\overline}
\newcommand{\dr}{\partial}
\newcommand{\ar}{\op\longrightarrow}

\newcommand{\ot}{\otimes}

\newcounter{theorem}
\newcounter{remark}
\newcounter{proposition}
\newcounter{lemma}
\newcounter{corollary}
\newcounter{definition}

\def\theremark{\arabic{remark}}

\def\thedefinition{\arabic{definition}}

\newenvironment{theo}{\refstepcounter{definition} \medskip
\noindent{\it Theorem \thedefinition.}}{\medskip}
\newenvironment{prop}{\refstepcounter{definition} \medskip
\noindent{\it Proposition \thedefinition.}}{\medskip}

\newcommand{\mar}[1]{}

\begin{document}
\hbox{}

{\parindent=0pt

{\large\bf Covariant Hamiltonian Field Theory. Path Integral
Quantization} 
\bigskip


{\bf D. Bashkirov\footnote{E-mail: bashkird@rol.ru} and
G. Sardanashvily}\footnote{E-mail: sard@grav.phys.msu.su;
Web: http://webcenter.ru/$\sim$sardan/}
\medskip

Department of Theoretical Physics,
Physics Faculty, Moscow State
University, 117234 Moscow, Russia

\bigskip

The Hamiltonian counterpart of classical Lagrangian field theory is
covariant Hamiltonian field theory where momenta correspond to
derivatives of fields with respect to all world coordinates. 
In particular, classical Lagrangian and covariant Hamiltonian
field theories are equivalent in the case of 
a hyperregular Lagrangian, and they are quasi-equivalent if a
Lagrangian is almost-regular. In order to quantize covariant
Hamiltonian field theory, one usually attempts to construct and quantize 
a multisymplectic generalization of the Poisson barcket. In the present
work, the path integral quantization of covariant Hamiltonian field
theory is suggested.
We use the fact that a covariant Hamiltonian field system is equivalent
to a certain Lagrangian system on a phase space which is quantized
in the framework of
perturbative quantum field theory. We show that, in the case of
almost-regular quadratic Lagrangians, path integral quantizations
of associated
Lagrangian and Hamiltonian field systems are equivalent.




}


\section{INTRODUCTION}

As is well-known, the familiar symplectic Hamiltonian technique applied
to field theory leads to instantaneous Hamiltonian formalism on an
infinite-dimensional phase space coordinated by field
functions at some instant of time (see (Gotay, 1991) for the strict
mathematical exposition of this formalism). 
The true Hamiltonian counterpart of classical first-order 
Lagrangian field theory is
covariant Hamiltonian formalism, where canonical momenta $p^\m_i$
correspond to derivatives $y^i_\m$ of fields $y^i$ with respect to all
world coordinates $x^\m$. This formalism has
been vigorously developed since the 1970s in its polysymplectic,
multisymplectic and Hamilton -- De Donder variants (see
(Giachetta et al., 1997, 1999;
Echeverr\'{\i}a-Enr\'{\i}quez et al., 2000;  
H\'elein and Kouneiher, 2002; Echeverr\'{\i}a-Enr\'{\i}quez et al., 
2004) and references 
therein). In order to quantize covariant
Hamiltonian field theory, one usually attempts to construct 
multisymplectic generalization of the Poisson barcket with respect to
the derivatives $\dr/\dr y^i$ and $\dr/\dr p^\m_i$ (Kanatchikov, 1999;
Castrill\'on L\`opez and Marsden, 2003; Forger et al., 2003). 

We have
suggested to quantize covariant (polysymplectic)
Hamiltonian field theory in path integral terms
(Sardanashvily, 1994). In the present work, this quantization scheme is
modified 
owing to the fact that a polysymplectic Hamiltonian system with a
Hamiltonian $\cH(x^\m,y^i,p^\m_i)$ is equivalent to a Lagrangian system
with the Lagrangian
\mar{m1}\beq
\cL_\cH (x^\m,y^i,p^\m_i,y^i_\la)=p_i^\la
y^i_\la-\cH(x^\m,y^i,p^\m_i,y^i_\la) \label{m1} 
\eeq
of the variables $y^i$ and $p^\m_i$. This Lagrangian system can be
quantized in the framework of 
familiar perturbative quantum field theory. If there is no constraint and 
the matrix $\dr^2\cH/\dr p^\m_i\dr p^\nu_j$ is nondegenerate and
positive-definite, this quantization is given by 
the generating functional 
\mar{m2}\beq
Z=\cN^{-1}\int\exp\{\int(\cL_\cH +\Lambda +
iJ_iy^i+iJ^i_\m p^\m_i) dx \}\op\prod_x [dp(x)][dy(x)] \label{m2}
\eeq
of Euclidean Green functions, where $\Lambda$ comes from the normalization
condition
\be
\int \exp\{\int(-\frac12\dr_\m^i\dr_\nu^j\cH p^\m_i
p^\nu_j+\La)dx\}\op\prod_x[dp(x)]=1. 
\ee
If
a Hamiltonian $\cH$ is degenerate, the Lagrangian
$\cL_\cH$ (\ref{m1}) may admit gauge symmetries. In this case,
integration of a generating functional along gauge group orbits must be
finite. If
there are constraints, the Lagrangian system with a Lagrangian 
$\cL_\cH$ (\ref{m1}) restricted to the constraint manifold is quantized. 

In order to verify this path quantization scheme, we apply it to Hamiltonian
field systems associated to Lagrangian field systems
with quadratic Lagrangians
\mar{N12}\beq
\cL=\frac12 a^{\la\m}_{ij} y^i_\la y^j_\m +
b^\la_i y^i_\la + c, \label{N12}
\eeq
where $a$, $b$ and $c$ are functions of world coordinates $x^\m$ and 
field variables $y^i$. Note that, in the framework of perturbative
quantum field theory, any Lagrangian is split into the sum of a
quadratic Lagrangian (\ref{N12}) and an interaction  
term quantized as a perturbation.

For instance, let the
Lagrangian (\ref{N12}) be hyperregular, i.e., the matrix function $a$
is nondegenerate. 
Then there exists a unique associated
Hamiltonian system whose Hamiltonian $\cH$ is quadratic in momenta
$p^\m_i$, and so is the Lagrangian $\cL_\cH$ (\ref{m1}). 
If the matrix function $a$ is positive-definite on an Euclidean space-time,
the generating functional (\ref{m2}) 
is a Gaussian integral 
of momenta $p^\m_i(x)$. Integrating $Z$ with respect to
$p^\m_i(x)$, one restarts the generating functional of quantum
field theory with the original Lagrangian $\cL$ (\ref{N12}). 
We extend this result to field theories 
with almost-regular 
Lagrangians $\cL$ (\ref{N12}), e.g., Yang--Mills gauge
theory. The key point is that, though such a Lagrangian
$\cL$ yields constraints and admits different 
associated Hamiltonians $\cH$, all the Lagrangians $\cL_\cH$
coincide on the constraint manifold, and we have a unique constrained
Hamiltonian system which is quasi-equivalent to the original Lagrangian one
(Giachetta at al., 1997, 1999). 

\section{COVARIANT HAMILTONIAN FIELD THEORY}

We follow the geometric formulation of classical field theory where
classical fields are represented by sections of fiber bundles.
Let $Y\to X$ be a smooth fiber bundle 
provided with bundle coordinates $(x^\m,y^i)$. The
configuration space of Lagrangian field theory on $Y$ is
the first-order jet manifold $J^1Y$ of $Y$. It is equipped with the
bundle coordinates
$(x^\m,y^i,y^i_\m)$ compatible with the composite fibration
\be
J^1Y\ar^{\pi^1_0} Y\ar^\pi X.
\ee
Any section $s$ of $Y\to X$ is prolonged to the section
$J^1s$ of $J^1Y\to X$ such that $y^i_\m\circ J^1s=\dr_\m s^i$.
A first-order Lagrangian is defined as a horizontal density
\mar{cmp1}\beq
L=\cL\om: J^1Y\to\op\w^nT^*X, \qquad \om=dx^1\w\cdots dx^n, \qquad n=\dim X,
\label{cmp1}
\eeq
on the jet manifold $J^1Y$. The corresponding 
Euler--Lagrange equations are given by the subset
\mar{b327}\beq
(\dr_i- d_\la\dr^\la_i)\cL=0,   \qquad
 d_\la=\dr_\la + y^i_\la\dr_i + y^i_{\la\m}\dr^\m_i, \label{b327}
\eeq
of the second-order jet manifold $J^2Y$ of $Y$ coordinated by
$(x^\m,y^i,y^i_\la, y^i_{\la\m})$. A section $s$ of $Y\to X$ is a
solution of these equations if its second jet prolongation $J^2s$ lives
in the subset (\ref{b327}). 

The phase space of covariant (polysymplectic) Hamiltonian field theory
on $Y$ is the Legendre bundle
\mar{00}\beq
\Pi=\op\w^nT^*X\op\ot_YV^*Y\op\ot_YTX=V^*Y\w(\op\w^{n-1}T^*X), \label{00}
\eeq
where $V^*Y$ is the vertical cotangent bundle of $Y\to X$. 
The Legendre bundle $\Pi$ is equipped with the holonomic bundle coordinates
$(x^\la,y^i,p^\m_i)$ compatible with the composite fibration 
\be
\Pi\ar^{\pi_Y} Y\ar^\pi X. 
\ee
It is endowed with the canonical polysymplectic form
\be
\Om =dp_i^\la\w dy^i\w \om\ot\dr_\la.
\ee
A covariant Hamiltonian $\cH$ on $\Pi$ (\ref{00}) is defined
as a section 
$p=-\cH$ of the trivial one-dimensional fiber bundle 
\mar{N41}\beq
Z_Y=T^*Y\w(\op\w^{n-1}T^*X)\to \Pi, \label{N41}
\eeq
equipped with holonomic bundle coordinates $(x^\la,y^i,p^\m_i,p)$.
This fiber bundle is provided with the canonical multisymplectic
Liouville form
\be
\Xi= p\om + p^\la_i dy^i\w\om_\la, \qquad \om_\la=\dr_\la\rfloor\om.
\ee
The pull-back of $\Xi$ onto $\Pi$ by a Hamiltonian $\cH$ is a Hamiltonian form
\mar{b418}\beq
 H=\cH^*\Xi_Y= p^\la_i dy^i\w \om_\la -\cH\om  \label{b418}
\eeq
 on $\Pi$. The corresponding covariant Hamilton equations on $\Pi$ 
are given by the closed submanifold
\mar{b4100}\beq
y^i_\la=\dr^i_\la\cH, \qquad  p^\la_{\la i}=-\dr_i\cH \label{b4100}
\eeq
of the jet manifold $J^1\Pi$ of $\Pi$. A section $r$ of $\Pi\to X$ is a
solution of these equations if its jet prolongation $J^1r$ lives in the
submanifold (\ref{b4100}).

\begin{prop} \label{m10} \mar{m10}
A section $r$ of $\Pi\to X$ is a
solution of the covariant Hamilton equations (\ref{b4100}) iff it
satisfies the condition
$r^*(u\rfloor dH)= 0$
for any vertical vector field $u$ on $\Pi\to X$.
\end{prop}

\begin{prop} \label{m11} \mar{m11}
A section $r$ of $\Pi\to X$ is a
solution of the covariant Hamilton equations (\ref{b4100}) iff
it is a solution of
the Euler--Lagrange equations for the first-order Lagrangian
\mar{m5}\beq
L_\cH=h_0(H)=\cL_\cH\om= (p^\la_i y^i_\la-\cH)\om \label{m5}
\eeq
on $J^1\Pi$, where $h_0$ sends exterior forms on $\Pi$ onto horizontal
exterior forms on $J^1\Pi\to X$ by the rule $h_0(dy^i)=y^i_\la dx^\la$. 
\end{prop}

Note that, for any section $r$ of $\Pi\to X$, the pull-backs
$r^*H$ and $J^1r^*L_\cH$ coincide. This fact and Proposition \ref{m11}
motivate us 
to quantize covariant Hamiltonian field theory with a Hamiltonian $\cH$
on $\Pi$ as a
Lagrangian system with the Lagrangian $L_\cH$ (\ref{m5}).

Furthermore, let $i_N:N\to \Pi$ be a closed imbedded subbundle of the Legendre
bundle $\Pi\to Y$ which is regarded as a constraint space of a
covariant Hamiltonian field system with a Hamiltonian $\cH$. This
Hamiltonian system is restricted to $N$ as follows. Let $H_N=i^*_NH$ be
the pull-back of the Hamiltonian form $H$ (\ref{b418}) onto $N$.
The constrained Hamiltonian form $H_N$ defines the
constrained Lagrangian 
\mar{cmp81}\beq
L_N=h_0(H_N)=(J^1i_N)^*L_\cH \label{cmp81}
\eeq
on the jet manifold $J^1N_L$ of the fiber bundle $N_L\to X$. 
The Euler--Lagrange equations for this Lagrangian are called the 
constrained Hamilton equations. 

Note that, in fact, the Lagrangian $L_\cH$ (\ref{m5})
is the pull-back onto $J^1\Pi$ of the horizontal form $L_\cH$ on
the bundle product $\Pi\op\times_Y J^1Y$ over $Y$ by the canonical map
$J^1\Pi\to \Pi\op\times_Y J^1Y$. Therefore, the constrained Lagrangian
$L_N$ (\ref{cmp81}) is simply the restriction of $L_\cH$ to
$N\op\times_Y J^1Y$. 

\begin{prop} \label{m12} \mar{m12}
A section $r$ of the fiber bundle $N\to X$ is a solution of constrained
Hamilton equations iff it satisfies the condition
$r^*(u_N\rfloor dH)= 0$
for any vertical vector field $u_N$ on $N\to X$.
\end{prop}

It follows from Proposition \ref{m10} and Proposition \ref{m12} that
any solution of the covariant Hamilton equations (\ref{b4100}) which
lives in the constraint manifold $N$ is also a solution of the
constrained Hamilton equations on $N$. This fact
motivates us
to quantize covariant Hamiltonian field theory on a constraint manifold
$N$ as a Lagrangian system with the pull-back Lagrangian $L_N$ (\ref{cmp81}).

Since a constraint manifold is assumed to be a closed imbedded
submanifold of $\Pi$, there exists its open neighbourhood $U$ which is
a fibered manifold $U\to N$. If $\Pi$ is a fibered manifold $\pi_N:\Pi\to N$ 
over $N$, it is often convenient to quantize a Lagrangian system on
$\Pi$ with the pull-back Lagrangian $\pi_N^*L_N$,
but integration of the corresponding generating functional along the
fibers of $\Pi\to N$ must be finite.  

In order to verify this quantization scheme, let us associate to a
Lagrangian field system on $Y$ a covarinat Hamiltonian system on $\Pi$,
then let us quantize this Hamiltonian system and compare this quantization
with that of an original Lagrangian system.

\section{ASSOCIATED LAGRANGIAN AND HAMILTONIAN SYSTEMS}

In order to relate classical Lagrangian and covariant Hamiltonian
field theories, let us recall that, besides the Euler--Lagrange
equations, a Lagrangian $L$ (\ref{cmp1}) also yields the Cartan
equations which are given by the subset 
\mar{b336}\ben
&& (\ol y^j_\m- y^j_\m)\dr^\la_i\dr^\m_j\cL=0, \qquad
\dr_i\cL-\ol d_\la\dr_i^\la\cL +(\ol y^j_\m- y^j_\m)\dr_i\dr^\m_j\cL=0,
\label{b336}\\
&& \ol d_\la=\dr_\la +\ol y^i_\la\dr_i +\ol y^i_{\la\m}\dr^\m_i, \nonumber
\een
of the repeated jet manifold $J^1J^1Y$ coordinated by
$(x^\m,y^i,y^i_\la,\ol y^i_\la,\ol y^i_{\la\m})$. A solution of the
Cartan equations is a section $\ol s$ of the jet bundle $J^1Y\to X$
whose jet prolongation $J^1\ol s$ lives in the subset (\ref{b336}).
Every solution $s$ of the Euler--Lagrange equations (\ref{b327})
defines the solution  $J^1s$ of the Cartan equations (\ref{b336}). If
$\ol s$ is a solution of the Cartan equations and $\ol s=J^1s$, then
$s$ is a solution of the Euler--Lagrange equations. If a Lagrangian $L$
is regular, the equations (\ref{b327}) and (\ref{b336}) are equivalent.

Any Lagrangian $L$ (\ref{cmp1}) yields the Legendre map
\mar{m3}\beq
\wh L: J^1Y\ar_Y \Pi, \qquad p^\la_i\circ\wh L=\dr^\la_i\cL, \label{m3}
\eeq
over $\id Y$ whose image $N_L=\wh L(J^1Y)$ is called the Lagrangian
constraint space. 
A Lagrangian $L$ is said to be hyperregular if the Legendre map
(\ref{m3}) is a 
diffeomorphism. A Lagrangian $L$ is called almost-regular if the
Lagrangian constraint space  is a closed imbedded subbundle 
$i_N:N_L\to \Pi$ of the
Legendre bundle $\Pi\to Y$ and the surjection 
$\wh L:J^1Y\to N_L$ 
is a submersion (i.e., a fibered manifold) whose fibers are connected. 
Conversely, any Hamiltonian $\cH$ yields 
the Hamiltonian map 
\mar{415}\beq
\wh H: \Pi\ar_Y J^1Y, \qquad y_\la^i\circ\wh H=\dr^i_\la\cH. \label{415}
\eeq

A Hamiltonian $\cH$ on $\Pi$ is said to be associated
to a Lagrangian $L$ on $J^1Y$ if $\cH$ satisfies the relations
\mar{2.30a,b}\ben
&&\wh L\circ\wh H\circ \wh L=\wh L, \qquad 
p^\m_i=\dr^\m_i\cL (x^\m,y^i,\dr^j_\la\cH), \qquad (x^\m,y^i,p^\m_i)\in N_L,
\label{2.30a} \\
&&\wh H^*L_\cH=\wh H^*L, \qquad
p^\m_i\dr^i_\m\cH-\cH=\cL(x^\m,y^j,\dr^j_\la\cH). \label{2.30b}
\een
If an associated Hamiltonian $\cH$ exists, 
the Lagrangian constraint space
$N_L$ is given by the coordinate relations (\ref{2.30a}) and
$\wh H\circ \wh L$ is a projector from $\Pi$ onto $N_L$.

For instance, any hyperregular Lagrangian $L$ admits a unique
associated Hamiltonian $\cH$ such that  
\be
\wh H=\wh L^{-1}, \qquad \cH=p^\m_i \wh L^{-1}{}_\m^i -\cL(x^\la, y^i,
\wh L^{-1}{}_\la^i). 
\ee
In this case, any solution $s$ of the Euler--Lagrange equations
(\ref{b327}) defines the solution $r=\wh L\circ J^1s$,
of the covariant Hamilton equations (\ref{b4100}).
Conversely, any solution $r$ of these Hamilton equations 
yields the solution $s=\pi_Y\circ r$ of the Euler--Lagrange equations 
(\ref{b327}).

A degenerate Lagrangian need not admit an associated Hamiltonian. If
such a Hamiltonian exists, it is not necessarily unique.
Let us restrict our
consideration to almost-regular Lagrangians. From the physical
viewpoint, the most of Lagrangian field theories is of this type.
From the mathematical one, this notion of degeneracy is particularly
appropriate for the study of relations between Lagrangian and covariant
Hamiltonian formalisms as follows. 

\begin{theo}\label{3.23} \mar{3.23}
Let $L$ be an almost-regular Lagrangian and $\cH$ an associated Hamiltonian.
Let a section $r$ of $\Pi\to X$
be a  solution of the covariant Hamilton equations (\ref{b4100})
for $\cH$. If $r$ lives in the constraint manifold $N_L$, then 
$s=\pi_Y\circ r$ satisfies the Euler--Lagrange
equations (\ref{b327}) for $L$, while $\ol s=\wh H\circ r$ obeys the
Cartan equations (\ref{b336}). Conversely, let $\ol s$ be a solution of the
Cartan equations (\ref{b336}) for $L$.
If $\cH$ satisfies
the relation
\be
\wh H\circ \wh L\circ \ol s=J^1(\pi^1_0\circ\ol s),
\ee
the section $r=\wh L\circ \ol s$
of the Legendre bundle $\Pi\to X$ is a solution of the
Hamilton equations (\ref{b4100}) for $\cH$. If $\ol s=J^1s$, we obtain
the relation between solutions the Euler--Lagrange equations and the
covariant Hamilton ones.
\end{theo}

By virtue of Theorem \ref{3.23}, one need a set of
different associated Hamiltonians in order to
recover all solutions of 
the Euler--Lagrange and Cartan equations for an almost-regular
Lagrangian $L$. We can overcome this ambiguity as follows.

\begin{prop} \label{m0} \mar{m0}
Let $\cH$, $\cH'$ be two different Hamiltonians associated to an
almost-regular Lagrangian $L$. Let $H$, $H'$ be the corresponding
Hamiltonian forms (\ref{b418}). Their pull-backs $i^*_NH$ and $i^*_NH'$
onto the Lagrangian constraint manifold $N_L$ coincide with each other.
\end{prop}

It follows that, if an almost-regular Lagrangian admits associated
Hamiltonians $\cH$, it defines a unique constrained
Hamiltonian form $H_N=i^*_NH$ on
the Lagrangian constraint manifold $N_L$ and a unique constrained
Lagrangian $L_N=h_0(H_N)$ (\ref{cmp81})
on the jet manifold $J^1N_L$ of the fiber bundle $N_L\to X$. 
For any Hamiltonian $\cH$ 
associated to $L$, every solution $r$
of the Hamilton equations which lives in the Lagrangian constraint space $N_L$
is a solution of the constrained Hamilton equations for $L_N$.

\begin{theo}\label{3.01} \mar{3.01} Let an almost-regular Lagrangian
$L$ admit associated Hamiltonians. A section
$\ol s$ of the jet bundle $J^1Y\to X$ is a solution of the Cartan
equations for $L$ iff 
$\wh L\circ \ol s$ is a solution of  the constrained Hamilton equations.
In particular, any solution $r$ of the constrained Hamilton equations
provides the solution $\ol s=\wh H\circ r$ of the Cartan equations.
\end{theo}

Theorem \ref{3.01}  shows that the constrained Hamilton equations and
the Cartan equations are quasi-equivalent. Thus, one can associate to an
almost-regular Lagrangian $L$ (\ref{cmp1}) a unique constrained
Lagrangian system on the
constraint Lagrangian manifold $N_L$ (\ref{2.30a}). Let us compare
quantizations of these Lagrangian systems on $Y$ and $N_L\subset \Pi$
in the case of an 
almost-regular quadratic Lagrangian $L$.

\section{QUADRATIC DEGENERATE SYSTEMS}

Given a fiber bundle $Y\to X$,
let us consider a  quadratic Lagrangian $L$ (\ref{N12}), 
where $a$, $b$ and $c$ are local functions on $Y$. This property is
coordinate-independent since $J^1Y\to Y$ is an affine bundle modelled
over the vector bundle $T^*X\op\ot_Y VY$, where $VY$ denotes the
vertical tangent bundle of $Y\to X$.
The associated Legendre map (\ref{m3}) reads
\mar{N13}\beq
p^\la_i\circ\wh L= a^{\la\m}_{ij} y^j_\m +b^\la_i. \label{N13}
\eeq

Let a Lagrangian $L$ (\ref{N12}) be almost-regular, i.e.,
the matrix function $a$ is a linear bundle morphism 
\mar{m38}\beq
a: T^*X\op\ot_Y VY\to \Pi, \qquad p^\la_i=a^{\la\m}_{ij} \ol y^j_\m,
\label{m38} 
\eeq
of constant rank, where $(x^\la,y^i,\ol y^i_\la)$ are bundle
coordinates on $T^*X\op\ot_Y VY$. Then
the Lagrangian constraint space $N_L$ 
(\ref{N13}) is an affine subbundle of the Legendre bundle $\Pi\to Y$.
Hence, $N_L\to Y$ has a global section. For the sake of simplicity, let us
assume that it is the canonical
zero section $\wh 0(Y)$ of $\Pi\to Y$. 
The kernel
of the Legendre map (\ref{N13})  is also an affine
subbundle of the affine jet bundle $J^1Y\to Y$. Therefore, it admits a
global section 
\mar{N16}\beq
\G: Y\to \Ker\wh L\subset J^1Y, \qquad
a^{\la\m}_{ij}\G^j_\m + b^\la_i =0,  \label{N16}
\eeq
which is a connection on $Y\to X$. If the Lagrangian (\ref{N12}) is regular, the
connection (\ref{N16}) is unique.

The forthcoming theorems are the key points of our analysis of
quadratic degenerate systems (Giachetta et al., 1997, 1999).

\begin{theo}\label{04.2}  There exists a linear bundle
morphism
\mar{N17}\beq
\si: \Pi\op\to_Y T^*X\op\otimes_YVY, \qquad
\ol y^i_\la\circ\si =\si^{ij}_{\la\m}p^\m_j, \label{N17}
\eeq
such that 
\mar{N45}\beq
a\circ\si\circ a=a, \qquad 
a^{\la\mu}_{ij}\si^{jk}_{\mu\al}a^{\al\nu}_{kb}=a^{\la\nu}_{ib}.
\label{N45}
\eeq
\end{theo}

Note that $\si$ is not unique, but it
falls into the sum $\si=\si_0+\si_1$ such that
\mar{N21}\beq
\si_0\circ a\circ \si_0=\si_0, \qquad a\circ\si_1=\si_1\circ a=0, \label{N21}
\eeq
where $\si_0$ is uniquely defined. For instance, there exists a
nondegenerate map $\si$ (\ref{N17}).

\begin{theo} \label{m15} \mar{m15} 
There are the splittings
\mar{N18,20}\ben
&& J^1Y=\cS(J^1Y)\op\oplus_Y \cF(J^1Y)=\Ker\wh L\op\oplus_Y{\rm Im}(\si_0\circ
\wh L), \label{N18} \\
&& y^i_\la=\cS^i_\la+\cF^i_\la= [y^i_\la
-\si_0{}^{ik}_{\la\al} (a^{\al\m}_{kj}y^j_\m + b^\al_k)]+
[\si_0{}^{ik}_{\la\al} (a^{\al\m}_{kj}y^j_\m + b^\al_k)], \nonumber\\
&& \Pi=\cR(\Pi)\op\oplus_Y\cP(\Pi)=\Ker\si_0 \op\oplus_Y N_L, \label{N20} \\
&& p^\la_i = \cR^\la_i+\cP^\la_i= [p^\la_i -
a^{\la\m}_{ij}\si_0{}^{jk}_{\m\al}p^\al_k] +
[a^{\la\m}_{ij}\si_0{}^{jk}_{\m\al}p^\al_k]. \nonumber
\een
\end{theo}

The relations (\ref{N21}) lead to the equalities
\mar{m25}\beq
a^{\la\m}_{ij}\cS^j_\m=0, \qquad \si_0{}^{jk}_{\m\al}\cR^\al_k=0,
\qquad \si_1{}^{jk}_{\m\al}\cP^\al_k=0, \qquad
\cR^\la_i\cF^i_\la=0. \label{m25}
\eeq
By virtue of these equalities, the Lagrangian (\ref{N12}) takes the form
\mar{cmp31}\beq
L=\cL\om, \qquad \cL=\frac12 a^{\la\m}_{ij}\cF^i_\la\cF^j_\m +c'. \label{cmp31}
\eeq
One can show that, this Lagrangian admits a set of associated Hamiltonians
\mar{N22}\beq
\cH_\G=(\cR^\la_i+\cP^\la_i)\G^i_\la
+\frac12
\si_0{}^{ij}_{\la\m}\cP^\la_i\cP^\m_j
+\frac12\si_1{}^{ij}_{\la\m}\cR^\la_i\cR^\m_j -c'\label{N22}
\eeq
indexed by connections $\G$ (\ref{N16}). Accordingly, the Lagrangian constraint
manifold (\ref{N13}) is given by the reducible constraints
\mar{bv1}\beq
\cR^\la_i=p^\la_i -
a^{\la\m}_{ij}\si_0{}^{jk}_{\m\al}p^\al_k=0. \label{bv1}
\eeq

Given a Hamiltonian $\cH_\G$, the corresponding Lagrangian 
(\ref{m5}) reads
\mar{m16}\beq
\cL_{\cH_\G}=\cR^\la_i(\cS^i_\la-\G^i_\la) +\cP^\la_i\cF_\la^i
-\frac12\si_0{}^{ij}_{\la\m}\cP^\la_i\cP^\m_j -
\frac12\si_1{}^{ij}_{\la\m} \cR^\la_i \cR^\m_j+ c'. \label{m16}
\eeq
Its restriction (\ref{cmp81}) to the Lagrangian constraint manifold $N_L$
(\ref{bv1}) is
\mar{bv2}\beq
L_N=\cL_N\om, \qquad \cL_N=\cP^\la_i\cF_\la^i
-\frac12\si_0{}^{ij}_{\la\m}\cP^\la_i\cP^\m_j + c'. \label{bv2}
\eeq
It is independent of the choice of a Hamiltonian (\ref{N22}).
Note that the Lagrangian $\cL_N$ may admit gauge symmetries due to the term
$\cP^\la_i\cF_\la^i$.

The Hamiltonian $\cH_\G$ yields the Hamiltonian map $\wh H_\G$
(\ref{415}) and the projector 
\mar{m30}\beq
T=\wh L\circ \wh H_\G,\qquad
p^\la_i\circ T=T^{\la j}_{i
\m}p^\m_j=a^{\la\nu}_{ik}\si_0{}^{kj}_{\nu\m} p^\m_j=\cP^\la_i,\label{m30}
\eeq
from $\Pi$ onto its
summand $N_L$ in the decomposition (\ref{N20}). It obeys the relations
\mar{m31}\beq
\si\circ T=\si_0, \qquad T\circ a=a. \label{m31}
\eeq
The projector $T$ (\ref{m30}) is a
linear morphism over $\id Y$. Therefore, $T:\Pi\to N_L$ is a vector bundle.
Let us consider the pull-back $L_\Pi=T^*L_N$ of the constrained
Lagrangian $L_N$ (\ref{bv2}) onto $\Pi$. By virtue of the relations
(\ref{m25}), it is given by the 
coordinate expression
\mar{m32}\beq
L_\Pi=\cL_\Pi\om, \qquad \cL_\Pi=p^\la_i\cF_\la^i
-\frac12\si_0{}^{ij}_{\la\m}p^\la_i p^\m_j + c'. \label{m32}
\eeq
This Lagrangian is gauge-invariant under the subgroup of the
gauge group of vertical automorphisms $\Phi$ of the affine bundle 
$\Pi\to Y$ such that $T\circ\Phi= T$. Clearly, this subgroup coincides
with the gauge group Aut$\,\Ker\si_0$ of vertical automorphisms of the
vector bundle $\Ker \si_0\to Y$.

In fact, the splittings (\ref{N18}) and (\ref{N20}) result from the
splitting of the vector bundle
\be
T^*X\op\ot_Y VY=\Ker a\op\oplus_Y E,
\ee
which can be provided with the adapted coordinates $(\ol y^a,\ol y^A)$ such that
$a$ (\ref{m38}) is brought into a diagonal matrix 
with nonvanishing
components $a_{AA}$. Then the Legendre bundle $\Pi\to Y$
(\ref{00}) is endowed with the dual (nonholonomic) coordinates $(p_a,p_A)$
where $p_A$ are coordinates on the Lagrangian constraint manifold $N_L$,
given by the irreducible constraints $p_a=0$.
Written relative to these coordinates, $\si_0$ becomes the
diagonal matrix 
\mar{m39}\beq
\si_0^{AA}=(a_{AA})^{-1}, \qquad \si_0^{aa}=0, \label{m39}
\eeq
while $\si_1^{Aa}=\si_1^{AB}=0$. Moreover, one can choose the 
coordinates 
$\ol y^a$ (accordingly, $p_a$) and the map $\si$ (\ref{N17}) such that 
$\si_1$ becomes a diagonal matrix with nonvanishing positive components 
$\si_1^{aa}=\cV^{-1}$,
where $\cV\om$ is a volume form on $X$. We further follow this choice
of the adapted coordinates $(p_a,p_A)$. Let us write
\mar{m41}\beq
p_a=M_a{}^i_\la p^\la_i, \qquad p_A=M_A{}^i_\la p^\la_i, \label{m41}
\eeq
where $M$ are the matrix functions on $Y$ obeying the relations
\be
M_a{}^i_\la a^{\la\m}_{ij}=0, \qquad M^{-1}{}_i^{\la a}\si_0{}^i_\la=0,
\qquad M_a{}^i_\la\cP^\la_i=0, \qquad M_A{}^i_\la\cR^\la_i=0. 
\ee
Then the Lagrangian $L_N$ (\ref{bv2}) 
with respect to the adapted coordinates $(p_a,p_A)$ takes the form
\mar{m20}\beq
\cL_N=M^{-1}{}_i^{\la A}p_A\cF_\la^i
-\frac12\op\sum_A(a_{AA})^{-1}(p_A)^2 + c', \label{m20}
\eeq

\section{QUANTIZATION}

Let us quantize a Lagrangian system with the Lagrangian $L_N$
(\ref{bv2}) on the constraint manifold $N_L$
(\ref{bv1}). 
In the framework of a perturbative quantum field theory,
we should assume that $X=\Bbb R^n$ and $Y\to X$ is a trivial affine bundle.
It follows that both the original coordinates $(x^\la,y^i,p^\la_i)$ and
the adapted coordinates $(x^\la,y^i,p_a,p_A)$ on the Legendre bundle
$\Pi$ are global. Passing to field theory on an Euclidean space
$\Bbb R^n$, we also assume that 
the matrix $a$ in the Lagrangian $L$ (\ref{cmp31}) is positive-definite,
i.e., $a_{AA}>0$.

Let us start from a Lagrangian (\ref{bv2}) without gauge symmetries.
Since the Lagrangian constraint space $N_L$ can be equipped with the 
adapted coordinates $p_A$, the generating functional of Euclidean Green
functions of the Lagrangian system in question reads 
\mar{m43}\beq
Z=\cN^{-1}\int\exp\{\int (\cL_N  +\frac12{\rm tr}\,\ln \ol \si_0+
iJ_iy^i+iJ^Ap_A)\om\}\op\prod_x 
[dp_A(x)][dy(x)], \label{m43}
\eeq
where $\cL_N$ is given by the expression (\ref{m20}) and
$\ol \si_0$ is the 
square matrix 
\be
\ol \si_0^{AB}=M^{-1}{}_i^{\la A} M^{-1}{}_j^{\m B} \si_0{}^{ij}_{\la\m}
=\dl^{AB}(a_{AA})^{-1}. 
\ee
The generating functional (\ref{m43}) a Gaussian
integral of variables $p_A(x)$. Its integration 
with respect to $p_A(x)$ under the condition $J^A=0$ restarts the generating
functional 
\mar{m44}\beq
Z=\cN^{-1}\int\exp\{\int (\cL + iJ_iy^i)\om\}\op\prod_x 
[dy(x)], \label{m44}
\eeq
of the original Lagrangian field system on $Y$ with the Lagrangian
(\ref{cmp31}). However, the generating functional (\ref{m43}) can not
be rewritten with 
respect to the original variables $p^\m_i$, unless $a$ is a
nondegenerate matrix function.

In order to overcome this difficulty, let us consider a Lagrangian
system on the whole Legendre manifold $\Pi$ with the Lagrangian $L_\Pi$
(\ref{m32}). 
Since this Lagrangian is constant along the fibers of the
vector bundle $\Pi\to N_L$, an integration of the generating functional
of this field model with respect to variables $p_a(x)$ should be finite.
One can choose the generating functional in the form
\mar{m21}\beq
Z=\cN^{-1}\int\exp\{\int (\cL_\Pi-\frac12\si_1{}^{ij}_{\la\m}p^\la_ip^\m_j
 +\frac12{\rm tr}\,\ln \si +
iJ_iy^i+iJ^i_\m p_i^\m)\om\} \op\prod_x 
[dp(x)][dy(x)]. \label{m21}
\eeq
Its integration with respect to momenta $p_i^\la(x)$ restarts the generating
functional (\ref{m44}) of the original Lagrangian system on $Y$. 
In order to obtain the generating functional (\ref{m21}), one can 
follow a procedure of quantization of gauge-invariant Lagrangian
systems. In the case of the Lagrangian $L_\Pi$ (\ref{m32}), this
procedure is rather trivial, since the space of momenta variables
$p_a(x)$ coincides with the translation subgroup of the gauge group
Aut$\,\Ker\si_0$.

Now let us suppose that the Lagrangian $L_N$ (\ref{bv2}) and, consequently,
the Lagrangian $L_\Pi$ (\ref{m32}) are invariant
under some gauge group $G_X$ of vertical automorphisms of the fiber bundle
$Y\to X$ (and the induced automorphisms of $\Pi\to X$) which acts freely
on the space of sections of $Y\to X$. 
Its infinitesimal
generators are represented by vertical vector fields
$u=u^i(x^\m,y^j)\dr_i$ on $Y\to X$ which give rise to the vector fields
\mar{m47}\beq
\ol u=u^i\dr_i-\dr_j u^ip^\la_i\dr^j_\la +d_\la u^i\dr_i^\la, \qquad
d_\la=\dr_\la +y_\la^i\dr_i, \label{m47}
\eeq
on $\Pi\op\times_Y J^1Y$. Let us also assume that $G_X$ is indexed
by $m$ parameter functions $\xi^r(x)$ such that
$u=u^i(x^\la,y^j,\xi^r)\dr_i$, where 
\mar{m48}\beq
u^i(x^\la,y^j,\xi^r)=u_r^i(x^\la,y^j)\xi^r
+u_r^{i\m}(x^\la,y^j)\dr_\m\xi^r \label{m48}
\eeq
are linear first order differential
operators on the space of parameters $\xi^r(x)$. The vector
fields $u(\xi^r)$ must satisfy the commutation relations
\be
[u(\xi^q),u(\xi'^p)]=u(c^r_{pq}\xi'^p\xi^q),
\ee
where $c^r_{pq}$ are structure constants.
The Lagrangian $L_\Pi$ (\ref{m32}) is invariant under the above mentioned
gauge transformations iff
its Lie derivative $\bL_{\ol u}L_\Pi$ along vector fields (\ref{m47})
vanishes, i.e.,
\mar{m49}\beq
(u^i\dr_i-\dr_j u^ip^\la_i\dr^j_\la +d_\la u^i\dr_i^\la)\cL_\Pi=0.
\label{m49} 
\eeq
Since the operator $\bL_{\ol u}$ is linear in momenta $p^\m_i$, the
condition (\ref{m49}) falls into the independent conditions
\mar{m50,1,2}\ben
&& (u^k\dr_k-\dr_j u^kp^\nu_k\dr^j_\nu +d_\nu
u^j\dr_j^\nu)(p^\la_i\cF_\la^i)=0,\label{m50}\\ 
&& (u^k\dr_k-\dr_j
u^kp^\nu_k\dr^j_\nu)(\si_0{}^{ij}_{\la\m}p^\la_ip^\m_j) =0, \label{m51}\\
&& u^i\dr_i c'=0. \label{m52}
\een
It follows that the Lagrangian $L_\Pi$ is gauge-invariant iff its three
summands are separately gauge-invariant. 

Note that, if the Lagrangian $L_\Pi$ on $\Pi$ is gauge-invariant, the
original Lagrangian $L$ (\ref{cmp31}) is also invariant under the
same gauge transformations. Indeed, one obtains at once from the
condition (\ref{m50}) that
\mar{m53}\beq
\ol u(\cF^i_\m)=\dr_ju^i\cF^j_\m,  \label{m53}
\eeq
i.e., the quantity $\cF$ is transformed as the dual of momenta $p$.
Then the condition 
(\ref{m51}) shows that the quantity $\si_0p$ is transformed by the same
law as $\cF$. It follows that the term $a\cF\cF$ in the Lagrangian 
$L$ (\ref{cmp31}) is transformed exactly as $a(\si_0 p)(\si_0
p)=\si_0pp$, i.e., is gauge-invariant. Then this Lagrangian is
gauge-invariant due to the equality (\ref{m52}). 

Since $\cS^i_\la=y^i_\la-\cF^i_\la$, one can easily derive from the formula
(\ref{m53}) the transformation law
\mar{m54}\beq
\ol u(\cS^i_\m)=d_\m u^i-\dr_ju^i\cF^j_\m=
d_\m u^i-\dr_ju^i(y^j_\m-\cS^j_\m) =\dr_\m u^i +\dr_ju^i\cS^j_\m
\label{m54} 
\eeq
of $\cS$. A glance at this expression shows that the gauge group $G_X$
acts freely on the space of sections $\cS(x)$ of the fiber bundle
$\Ker\wh L\to Y$ in the splitting (\ref{N18}). Let the number $m$ of
parameters of the gauge group $G_X$ do not exceed the fiber dimension
of $\Ker\wh L\to Y$. Then some combinations $b^r{}_i^\m\cS^i_\m$ of
$\cS^i_\m$ can be used as the gauge condition
\be
b^r{}_i^\m\cS^i_\m(x)-\al^r(x)=0, 
\ee
similar to the generalized Lorentz gauge in Yang--Mills gauge theory.

Turn now to quantization of a Lagrangian system with the
gauge-invariant Lagrangian $L_\Pi$ (\ref{m32}). In accordance with the
well-known quantization procedure, let us modify the generating
functional (\ref{m21}) as follows
\mar{m55}\ben
&& Z=\cN^{-1}\int\exp\{\int (\cL_\Pi-\frac12\si_1{}^{ij}_{\la\m}p^\la_ip^\m_j
 +\frac12{\rm tr}\,\ln \si -\frac12 h_{rs}\al^r\al^s +
iJ_iy^i+iJ^i_\m p_i^\m)\om\} \nonumber \\
&& \qquad \Delta\op\prod_x
\op\times^r\dl(b^r{}_i^\m\cS^i_\m(x)-\al^r(x))[d\al(x)] [dp(x)][dy(x)]=
\nonumber\\
&& \cN'^{-1}\int\exp\{\int (\cL_\Pi-\frac12\si_1{}^{ij}_{\la\m}p^\la_ip^\m_j
 +\frac12{\rm tr}\,\ln \si -\frac12 h_{rs}b^r{}_i^\m b^s{}_j^\la\cS^i_\m
\cS^j_\la +
iJ_iy^i+iJ^i_\m p_i^\m)\om\} \nonumber \\
&& \qquad \Delta\op\prod_x [dp(x)][dy(x)],
\label{m55}
\een
where 
\be
\int\exp\{\int (-\frac12 h_{rs}\al^r\al^s)\om\} \op\prod_x [d\al(x)]
\ee
is a Gaussian integral, and the factor $\Delta$ is defined by the condition
\be
\Delta\int \op\prod_x
\op\times^r\dl(u(\xi)(b^r{}_i^\m\cS^i_\m))[d\xi(x)]=1.
\ee
We have the linear second order differential operator 
\mar{m56}\beq
M^r_s\xi^s= u(\xi)(b^r{}_i^\m\cS^i_\m(x))=b^r{}_i^\m(\dr_\m u^i(\xi) +
\dr_ju^i(\xi)\cS^j_\m) \label{m56}
\eeq
on the parameter functions $\xi(x)$, and obtain $\Delta=\det M$.
Then the generating functional (\ref{m55}) takes the form
\mar{m57}\ben
&& Z=\cN'^{-1}\int\exp\{\int (\cL_\Pi-\frac12\si_1{}^{ij}_{\la\m}p^\la_ip^\m_j
 +\frac12{\rm tr}\,\ln \si -\frac12 h_{rs}b^r{}_i^\m b^s{}_j^\la\cS^i_\m
\cS^j_\la -\ol c_r M^r_sc^s + \nonumber \\
&& \qquad iJ_iy^i+iJ^i_\m p_i^\m)\om\} \op\prod_x [d\ol c][d c][dp(x)][dy(x)],
\label{m57}
\een
where $\ol c_r$, $c^s$ are odd ghost fields. Integrating $Z$
(\ref{m57}) with respect to momenta under the condition $J^i_\m=0$, we
come to the generating functional
\mar{m58}\beq
Z=\cN'^{-1}\int\exp\{\int (\cL-\frac12 h_{rs}b^r{}_i^\m b^s{}_j^\la\cS^i_\m
\cS^j_\la -\ol c_r M^r_sc^s+iJ_iy^i)\om\} \op\prod_x [d\ol c][d c][dy(x)]
\label{m58}
\eeq
of the original field model on $Y$ with the gauge-invariant Lagrangian
$L$ (\ref{cmp31}). 

Note that the Lagrangian
\mar{m59}\beq
\cL'=\cL-\frac12 h_{rs}b^r{}_i^\m b^s{}_j^\la\cS^i_\m
\cS^j_\la -\ol c_r M^r_sc^s \label{m59}
\eeq
fails to be gauge-invariant, but it admits the BRST symmetry whose
odd operator reads
\mar{m60}\ben
&& \vt=u^i(x^\m,y^i,c^s)\dr_i +d_\la u^i(x^\m,y^i,c^s)\dr_i^\la +\ol
v_r(x^\m,y^i,y^i_\m)\frac{\dr}{\dr\ol c_r} +  \nonumber\\
&& \qquad v^r(x^\m,y^i,c^s)
\frac{\dr}{\dr c^r}  +d_\la v^r(x^\m,y^i,c^s) \frac{\dr}{\dr c^r_\la}
+d_\m d_\la v^r(x^\m,y^i,c^s) \frac{\dr}{\dr c^r_{\m\la}}, 
 \label{m60}\\
&& d_\la=\dr_\la +y^i_\la\dr_i + y^i_{\la\m}\dr_i^\m +
c^r_\la \frac{\dr}{\dr c^r} +c_{\la\m}^r\frac{\dr}{\dr c^r_\m}. \nonumber
\een
Its components $u^i(x^\m,y^i,c^s)$ are given by the expression (\ref{m48}) 
where parameter functions $\xi^r(x)$ are replaced with the ghosts
$c^r$. The components $\ol v_r$ and $v^r$ of the BRST operator $\vt$
can be derived from the condition 
\be
\vt(\cL')= - h_{rs} M^r_qb^s{}_j^\la \cS^j_\la c^q -\ol v_rM^r_q c^q
+\ol c_r \vt(\vt(b^r{}_j^\la \cS^j_\la))=0
\ee
of the BRST invariance of $\cL'$. This condition falls into the two
independent relations 
\be
&&  h_{rs} M^r_qb^s{}_j^\la \cS^j_\la +\ol v_rM^r_q =0,\\
&& \vt(c^q)(\vt(c^p)(b^r{}_j^\la \cS^j_\la)) =u(c^p)(u(c^q)
(b^r{}_j^\la \cS^j_\la)) +u(v^r)(b^r{}_j^\la \cS^j_\la)= \\
&& \qquad u(\frac12c^r_{pq}c^pc^q+v^r)(b^r{}_j^\la \cS^j_\la)=0.
\ee
Hence, we obtain
\mar{m61}\beq
\ol v_r=-h_{rs}b^s{}_j^\la \cS^j_\la, \qquad v^r=-\frac12 c^r_{pq}c^pc^q.
\eeq

\section{HAMILTONIAN GAUGE THEORY}

For example, let us consider gauge theory of principal connections on
a principal 
bundle $P\to X$ with a structure Lie group $G$. 
Principal connections on $P\to X$ are
represented by sections of  the affine bundle
\mar{br3}\beq
C=J^1P/G\to X, \label{br3}
\eeq
modelled over the vector bundle $T^*X\ot V_GP$ \cite{book}. Here,
$V_GP=VP/G$ is the fiber bundle in Lie algebras $\cG$ of the group $G$.
Given the basis $\{\ve_r\}$ for $\cG$, we obtain the local fiber bases
$\{e_r\}$ for $V_GP$. The connection bundle $C$ (\ref{br3}) is
coordinated by $(x^\m,a^r_\m)$ such that, written
relative to these coordinates, sections 
$A=A^r_\m dx^\m\ot e_r$ of $C\to X$ are the familiar
local connection one-forms, regarded as gauge potentials. 

There is one-to-one correspondence between the
sections $\xi=\xi^r e_r$ of $V_GP\to X$ and the vector fields on $P$ which
are infinitesimal generators of one-parameter groups of vertical
automorphisms (gauge
transformations) of $P$.
Any section $\xi$ of $V_GP\to X$ yields the vector field 
\mar{br6}\beq
u(\xi)=u^r_\m \frac{\dr}{\dr a^r_\m}=(c^r_{pq}a^p_\m\xi^q+ \dr_\m\xi^r)
\frac{\dr}{\dr a^r_\m} \label{br6}
\eeq
on $C$, where $c^r_{pq}$ are the structure constants of the Lie algebra
$\cG$. 

The configuration space of gauge theory is the jet manifold 
$J^1C$ equipped with the coordinates $(x^\la,a^m_\la,a^m_{\m\la})$.
It admits the canonical splitting
(\ref{N18}) given by the coordinate expression
\mar{N31'}\beq
a^r_{\m\la}=\cS^r_{\m\la}+\cF^r_{\m\la}=
\frac12(a^r_{\m\la}+a^r_{\la\m}-c^r_{pq}a^p_\m 
a^q_\la) +\frac12(a^r_{\m\la}-a^r_{\la\m} +c^r_{pq}a^p_\m a^q_\la), 
\eeq
where $\cF$  is the strength of gauge fields
up to the factor 1/2.
The Yang--Mills Lagrangian on the configuration space $J^1C$
reads
\mar{5.1'}\beq
L_{\rm YM}=a^G_{pq}g^{\la\m}g^{\bt\n}\cF^p_{\la
\beta}\cF^q_{\m\n}\sqrt{\nm g}\,\om, \qquad  g=\det(g_{\m\nu}), \label{5.1'}
\eeq
where  $a^G$ is a non-degenerate $G$-invariant metric
in the dual of the Lie algebra of ${\got g}$ and $g$ is a pseudo-Riemannian
metric on $X$. 

The phase space $\Pi$ (\ref{00}) of the gauge theory is
endowed with the canonical coordinates
$(x^\la,a^p_\la,p^{\mu\la}_q)$. It admits the canonical splitting
(\ref{N20}) given by the coordinate expression
\mar{N32}\beq
p^{\mu\la}_m= \cR^{\mu\la}_m + \cP^{\mu\la}_m=
p^{(\mu\la)}_m + p^{[\mu\la]}_m=\frac{1}{2}(p^{\mu\la}_m+
p^{\la\mu}_m) + \frac{1}{2}(p^{\mu\la}_m-
p^{\la\mu}_m). \label{N32}
\eeq
With respect to this splitting, the Legendre map induced by the
Lagrangian (\ref{5.1'}) takes the form
\mar{5.2a,b} \ben
 &&p^{(\mu\la)}_m\circ\wh L_{YM}=0, \label{5.2a}\\
&&p^{[\mu\la]}_m\circ\wh
L_{YM}=4a^G_{mn}g^{\m\al}g^{\la\bt}
\cF^n_{\al\bt}\sqrt{|g|}. \label{5.2b}
\een
The equalities (\ref{5.2a}) define the Lagrangian constraint space
$N_L$ of Hamiltonian gauge theory. 
Obviously, it is an imbedded submanifold of $\Pi$, and the Lagrangian
$L_{\rm YM}$ is almost-regular.

In order to construct an associated Hamiltonian, let us consider
a connection $\G$ (\ref{N16}) on the fiber bundle $C\to X$ which
take their values into $\Ker\wh L$, i.e.,
\be
\G^r_{\la\m}-\G^r_{\m\la}+c^r_{pq}a^p_\la a^q_\m=0. 
\ee
Given a symmetric linear connection $K$ 
on $X$ and a principal connection $B$ on
$P\to X$, this connection reads
\be
\G^r_{\la\m}=\frac{1}{2} [\dr_\mu B^r_\la+\dr_\la B^r_\mu
-c^r_{pq}a^p_\la a^q_\mu  +  c^r_{pq}
(a^p_\la B^q_\m +a^p_\m B^q_\la)] -
K_\la{}^\bt{}_\mu(a^r_\bt-B^r_\bt). 
\ee
The corresponding 
Hamiltonian (\ref{N22}) associated to $L_{\rm YM}$ is
\be
\cH_\G=p^{\la\m}_r\G^r_{\la\m}+a^{mn}_Gg_{\mu\nu}
g_{\la\beta} p^{[\mu\la]}_m p^{[\nu\bt]}_n\sqrt{|g|}.
\ee
Then we obtain the Lagrangian 
\be
\cL_N=p^{[\la\m]}_r\cF^r_{\la\m}- a^{mn}_Gg_{\mu\nu}
g_{\la\beta} p^{[\mu\la]}_m p^{[\nu\bt]}_n\sqrt{|g|}
\ee
(\ref{bv2}) on the Lagrangian constraint manifold (\ref{5.2a}) and its
pull-back  
\mar{m63}\beq
L_\Pi=\cL_\Pi\om, \qquad \cL_\Pi=p^{\la\m}_r\cF^r_{\la\m}- a^{mn}_Gg_{\mu\nu}
g_{\la\beta} p^{[\mu\la]}_m p^{[\nu\bt]}_n\sqrt{|g|}, \label{m63}
\eeq
(\ref{m32}) onto $\Pi$. 

Both the Lagrangian $L_{\rm YM}$ (\ref{5.1'}) on $C$ and the Lagrangian 
$L_\Pi$ (\ref{m63}) on $\Pi$ are invariant under gauge transformations
whose infinitesimal generators are the lifts
\be
&& J^1u(\xi)=(c^r_{pq}a^p_\m\xi^q+\dr_\m\xi^r)
\frac{\dr}{\dr a^r_\m} + (c^r_{pq}(a^p_{\la\m}\xi^q +a^p_\m\dr_\la\xi^q)
+\dr_\la\dr_\m\xi^r)\frac{\dr}{\dr a^r_{\la\m}},\\
&& \ol u(\xi)=J^1u(\xi) -
c^r_{pq}p_r^{\la\m}\xi^q \frac{\dr}{\dr p_p^{\la\m}}
\ee
of the vector fields (\ref{br6}) onto $J^1C$ and $\Pi\op\times_C J^1C$,
respectively. We have the transformation laws
\be
J^1u(\xi)(\cF^r_{\la\m})=c^r_{pq}\cF^p_{\la\m}\xi^q, \qquad 
J^1u(\xi)(\cS^r_{\la\m})=
c^r_{pq}\cS^p_{\la\m}\xi^q +c^r_{pq}a^p_\m\dr_\la\xi^q+ \dr_\la\dr_\m\xi^r.
\ee
Therefore, one can choose the gauge conditions
\be
g^{\la\m}S^r_{\la\m}(x)-\al^r(x)= \frac12g^{\la\m}(\dr_\la a^r_\m(x)
+\dr_\m a^r_\la(x))-\al^r(x)=0,
\ee
which are the familiar generalized Lorentz gauge. The corresponding
second-order differential operator (\ref{m56}) reads
\be
M^r_s\xi^s=g^{\la\m}(\frac12 c^r_{pq} (\dr_\la a^r_\m
+\dr_\m a^r_\la)\xi^q +c^r_{pq}a^p_\m\dr_\la\xi^q +\dr_\la\dr_\m\xi^r).
\ee
Passing to the Euclidean space and repeating the quantization procedure
in Section 5, we come to the generating functional
\be
&& Z=\cN^{-1}\int\exp\{\int (p^{\la\m}_r\cF^r_{\la\m}- a^{mn}_Gg_{\mu\nu}
g_{\la\beta} p^{\mu\la}_m p^{\nu\bt}_n\sqrt{|g|}-\\
&& \qquad \frac18 a^G_{rs}g^{\al\nu}g^{\la\m}(\dr_\al a^r_\nu +\dr_\nu a^r_\al)
(\dr_\la a^s_\m +\dr_\m a^s_\la)
-g^{\la\m}\ol c_r(\frac12 c^r_{pq} (\dr_\la a^r_\m
+\dr_\m a^r_\la)c^q+c^r_{pq}a^p_\m c^q_\la +c_{\la\m}^r)  \\
&& \qquad + iJ_r^\mu a^r_\mu +iJ^r_{\m\la} p_r^{\m\la})\om\} \op\prod_x
[d\ol c][d c][dp(x)][da(x)].
\ee
Its integration with respect to momenta restarts the familiar
generating functional of gauge theory.

\end{document}